\documentclass[pre,twocolumn,floats,showpacs,superscriptaddress]{revtex4}
\usepackage{epsfig}
\usepackage{amsmath,amssymb,amsthm}
\usepackage{graphicx}
\usepackage{psfrag}
\usepackage{bm}
\def\n{\langle n \rangle}
\def\h{\langle h \rangle}

\def\s2{\sigma^2}

\begin{document}

\title{Critical wetting of a class of nonequilibrium interfaces: A computer simulation study}

\author{Elvira Romera}
\affiliation{Departamento de F\'isica At\'omica, Molecular y Nuclear, 
Universidad de Granada, Fuentenueva s/n, 18071 Granada, Spain}
\affiliation{Instituto Carlos I de F{\'\i}sica Te\'orica y
Computacional, Universidad de Granada, Fuentenueva s/n, 18071 Granada,
Spain}

\author{Francisco de los Santos}
\affiliation{Instituto Carlos I de F{\'\i}sica Te\'orica y
Computacional, Universidad de Granada, Fuentenueva s/n, 18071 Granada,
Spain}
\affiliation{
Departamento de Electromagnetismo y F{\'\i}sica de la
Materia, Universidad de Granada, Fuentenueva s/n, 18071 Granada,
Spain}

\author{Omar Al Hammal}
\affiliation{Instituto Carlos I de F{\'\i}sica Te\'orica y
Computacional, Universidad de Granada, Fuentenueva s/n, 18071 Granada,
Spain}
\affiliation{
Departamento de Electromagnetismo y F{\'\i}sica de la
Materia, Universidad de Granada, Fuentenueva s/n, 18071 Granada,
Spain}

\author{Miguel A. Mu\~noz}
\affiliation{Instituto Carlos I de F{\'\i}sica Te\'orica y
Computacional, Universidad de Granada, Fuentenueva s/n, 18071 Granada,
Spain}
\affiliation{
Departamento de Electromagnetismo y F{\'\i}sica de la
Materia, Universidad de Granada, Fuentenueva s/n, 18071 Granada,
Spain}

\date{\today}

\begin{abstract}

Critical wetting transitions under nonequilibrium conditions
are studied numerically and analytically by means of an interface-displacement
model defined by a Kardar-Parisi-Zhang equation, plus some extra terms
representing a limiting, short-ranged attractive wall.
Its critical behavior is characterized
in detail by providing a set of exponents for both the average height
and the surface order-parameter in one dimension. The emerging picture
is qualitatively and quantitatively different from recently reported
mean-field predictions for the same problem. Evidence is shown that
the presence of the attractive wall induces an anomalous scaling of
the interface local slopes.

\end{abstract}
\maketitle

\section{INTRODUCTION}

\begin{figure}[top]
\centerline{\psfig{figure=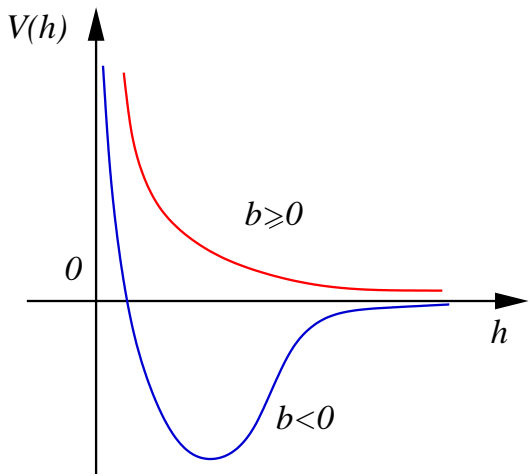,width=7.0cm}}
\caption{(Color online) Mean-field binding potential for positive and negative 
values of $b$. For $b<0$ the potential is attractive, exhibiting 
a well near the wall.}
\label{pot}
\end{figure}

Much scientific effort has gone into the study of equilibrium wetting
since, in the late seventies, Cahn introduced the idea that it can be
described as a phase transition \cite{cahn}. Among the various
theoretical approaches developed, interface displacement models have
proved particularly useful \cite{reviews_wetting}. Within this
perspective, the focus is on the interface that separates two
coexisting (bulk) phases confined by a wall or substrate, and the
wetting transition corresponds to the unbinding of the interface from
the wall. This happens upon a rise of the temperature when the wall
adsorbs preferentially one of the phases leading to a divergence of
the thickness of the adsorbed layer. The dynamics of such an interface
can be described at a coarse-grained level by the following continuum
stochastic growth equation
\cite{lipowsky}

\begin{equation}
\frac{\partial h({\bf x},t)}{\partial t}= D\nabla^2 h -\frac{\delta
V(h)}{\delta h} +\eta({\bf x},t).
\label{ew_wall}
\end{equation}
Here, $h({\bf x},t)$ is the local height of the interface from the wall, the
regions $y>h({\bf x})$ and $y<h({\bf x})$ corresponding to the two bulk phases. $D$
is the interfacial tension coefficient and $\eta$ is a Gaussian white noise
with zero mean and variance $\langle \eta({\bf x},t) \eta({\bf x}',t')
\rangle=2\sigma \delta({\bf x-x}')\delta(t-t')$ that mimics thermal
fluctuations. $V(h)$ accounts for the net interaction between the wall
and the interface and its form depends on the nature of the forces
between the particles in the bulk phases and with the wall, its
rigorous derivation from microscopic Hamiltonians being far from
trivial. If all the interactions are short-ranged, one may take in the
limit of large $h$ at phase coexistence
\cite{reviews_wetting}

\begin{equation}
V(h)=\int dx \left[b(T)e^{-h(x)} +\frac{c}{2} e^{-2h(x)}\right],
\end{equation}
where $T$ is the temperature \cite{parry}. The amplitude $c>0$ is a
repulsion whereas $b(T)$ vanishes linearly with the mean-field wetting
temperature, $T_w$, as $T-T_w$, and can represent either an effective
repulsion or attraction between the interface and the wall (see
Fig. \ref{pot}). At sufficiently low temperatures, $b<0$, the
equilibrium thickness of the wetting layer as given by the stationary
configurational average $\h$ is finite (pinned interface). This
corresponds to an attractive potential (see Fig. \ref{pot}). As the
temperature is raised the potential becomes less attractive and
eventually, above a certain value $b=b_w$, it no longer binds the
interface and $\h$ diverges. Within mean-field approximation, ignoring
spatial correlations, $\h$ follows from  $\partial
V(h)/\partial h=0$, whereby one finds for an attractive wall ($b<0$)
$\langle h \rangle =\ln(-c/b)$ and, consequently, a {\em critical
wetting} transition takes place as $b \to b_w=0$. Recently, effective
short-ranged, equilibrium critical wetting showing mean-field like
exponents seems to have been experimentally observed \cite{obsvcw}.

Extensions of equilibrium, interface displacement models to nonequilibrium
conditions have only been recently addressed and constitute a topic of 
ongoing research activity  \cite{noneq-wett,grinstein}. Supplementing equation (\ref{ew_wall}) with the
most relevant nonequilibrium nonlinear term $\lambda (\nabla h)^2$ \cite{kpz},
leads to a natural generalization of equation (\ref{ew_wall})
that assumes that the velocity of the interface depends on its local-slope,
\begin{equation}
\frac{\partial h(x,t)}{\partial t} =D\nabla^2 h +\lambda(\nabla h)^2+
be^{-h}+ce^{-2h} +\eta(x,t),
\label{kpz_wall}
\end{equation}
which is a Kardar-Parisi-Zhang (KPZ) interface \cite{kpz} interacting
via a short-ranged potential with a wall.  KPZ interfaces have a
nonzero average velocity, $v=\lambda \langle (\nabla h)^2 \rangle$,
and hence steady-state interfaces move on average thereby favoring one
of the phases over the other. Wetting, on the other hand, by
definition occurs at coexistence, i.e. zero average velocity of
the free (no wall) interface. This agrees with the thermodynamic
picture that at bulk coexistence any arbitrary fraction of the system
may be in one phase, with the remainder in the other.  Therefore, a
constant $a_c=-v$ needs to be included in equation (\ref{kpz_wall}) to
study wetting transitions driven by the wall. The nonequilibrium
analog of equilibrium critical wetting corresponds to the depinning
transition at $a_c$ as $b \to b_w^-$.  Clearly, for $\lambda=0$ and
$a_c=0$, the model reduces to the equilibrium one. In equilibrium, a
constant force term, $a$, in the interfacial equation measures the
deviation from bulk coexistence (i.e. it represents the chemical
potential difference between the two phases) while here it plays a
similar role by balancing the force exerted by the KPZ nonlinearity on
the tilted parts of the interface, thereby guaranteeing that the
average velocity of the free interface is zero.

Owing to the lack of $h \leftrightarrow -h$ symmetry of the KPZ
dynamics, it is necessary to specify either the relative position of
the wall with respect to the interface, upper or lower, for a fixed
sign of $\lambda$, or reversely, the sign of $\lambda$ after a wall
position has been arbitrarily chosen.  
In earlier studies of nonequilibrium (complete) wetting (see below),
these two different physical situations lead to the existence of
by-now-well-documented two different universality classes, called {\it
multiplicative noise 1} (MN1) and {\it multiplicative noise 2} (MN2)
respectively \cite{reviews}. Here, we take $\lambda <0$ in
Eq. (\ref{kpz_wall}) (i.e. the critical wetting counterpart of MN1)
which has shown to have a much richer phenomenology than that of
$\lambda >0$ in other studies of nonequilibrium wetting
\cite{reviews,cw_meanfield}.  The analysis of the case $\lambda>0$ will be tackled
elsewhere.

\begin{figure}[top]
\centerline{\psfig{figure=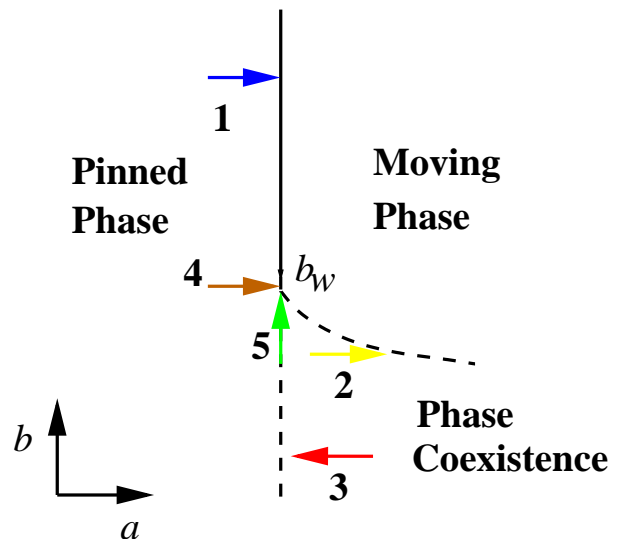,width=8.0cm}}
\caption{
(Color online)
Schematic phase diagram for $\lambda <0$ and a lower wall in the
$a-b$ plane (Eq. \ref{kpz_wall2}). The vertical line corresponds to the critical value $a=a_c$,
and the arrows denote the different types of transitions explained in
the text. 
Path 1: complete wetting (upon approaching $a_c$). 
Path 2: non-trivial depinning transition at $a^*(b)$. 
Path 3: First-order pinning transition. 
Path 4: Multicritical complete wetting. 
Path 5: Critical wetting.}
\label{phase_diagram}
\end{figure}

Thus defined, this model system, arguably the simplest nonequilibrium
one, has served for the study of universality issues in nonequilibrium
wetting. In particular, by fixing $b>b_w$, i.e. in the presence of a
repulsive wall, and letting $a\to a_c^-$ (path 1 in figure
\ref{phase_diagram}) nonequilibrium {\em complete wetting} transitions
(MN1) were investigated. For $b<b_w$ (attractive wall) and varying $a$
there is a rich phenomenology: the pinned and the
depinned phases lose their stability at different values of $a$,
giving rise to a continuous depinning transition at $a^*(b)>a$ in
the directed-percolation universality class (see
Fig. \ref{phase_diagram}, path 2) \cite{Romu}, and a first-order
phase-transition along path 3, at $a=a_c$.  In the broad interval
$a_c<a<a^*(b)$ both phases coexist (see \cite{reviews} and references
therein).
Tricritical behavior along path 4 was analyzed by Ginelli et
al. \cite{ginelli}, and a preliminary study of nonequilibrium critical
wetting (path 5) was presented in \cite{nos}.

In this paper we investigate numerically and analytically nonequilibrium
critical wetting  (path 5 in figure \ref{phase_diagram}) as defined by
equation (\ref{kpz_wall}). The focus will be on one-dimensional systems only.
Higher system dimensionalities were studied in \cite{cw_meanfield} by
a mean-field analytic approximation to Eq. (\ref{kpz_wall}) which
revealed the existence of three different regimes of scaling
behavior. Their connection with our findings (or the lack of it) is
discussed in the last section.

\section{MODELS AND OBSERVABLES}

Our continuous model is defined by the stochastic growth equation
\begin{equation}
\frac{\partial h(x,t)}{\partial t} =D\nabla^2 h +\lambda(\nabla h)^2+a_c +
be^{-h}+ce^{-2h} +\eta(x,t),
\label{kpz_wall2}
\end{equation}
with, as explained above, $a_c=-\lambda \langle (\nabla h)^2 \rangle$.
At the critical wetting transition, i.e.  as $b$ approaches $b_w$ from
below, the average stationary thickness of the wetting layer diverges
continuously as $\h \sim |b-b_w|^{\beta_h}$, where $\beta_h$ is a
critical exponent. At $b=b_w$, $\langle h(t) \rangle \sim
t^{\theta_h}$ for asymptotically long times. Two other exponents we
study are the dynamic and the correlation length exponents, $z$ and
$\nu$ respectively, defined through their usual expressions $\xi(t)
\sim t^{1/z}, \xi \sim |b-b_w|^{-\nu}$, where $\xi$ is the correlation length.
They are related to the previous ones by the scaling form
$\theta_h=\beta_h/z\nu$.

Some of these exponents may be written in terms of known KPZ
exponents. In particular, since at $b=b_w$ the interface is
asymptotically free, the dynamic exponent retains its one-dimensional
free KPZ value $z=3/2$ \cite{grinstein}. 

Also, as we illustrate now, $\theta_h$ is given by the exponent
characterizing the growth of the interfacial width in the KPZ, $W(t) \sim
t^{\theta_W}$, and therefore, $\theta_h=\theta_W =1/3$ in one
dimension. This can be understood as follows: interfacial fluctuations
are cutoff owing to the presence of the wall, as a result of which
there is an effective fluctuation-induced repulsion between the wall
and the interface. The latter can be estimated by noting that the wall
makes itself felt when the mean interfacial separation $\h$ is of the
same order as the average extent of the interface fluctuations,
$\delta h$. From $\delta h \sim \xi^{\zeta}$, where $\zeta$ is the
usual KPZ roughness exponent, and the definition of $\nu$ we find that
the effective repulsion force has the form $h^{-1/\zeta \nu}$ and is
therefore long-ranged in $d=1$ where $\zeta=1/2>0$. Comparing this
force with the deterministic one in the Langevin equation
(\ref{kpz_wall2}), which is short-ranged, it is straightforward to
conclude that in $d=1$ fluctuations dominate the unbinding of the
interface. As fluctuations are governed, in the regime where the
interface is asymptotically free, by the growth exponent of the KPZ,
then $\theta_h=1/3$.  This result,
has been verified in our computer simulations (see below). 
Note that this, as well as any other exponent
computed exactly at the critical point, is path independent and,
therefore, holds also for complete wetting.

\begin{figure}[top]
\centerline{\psfig{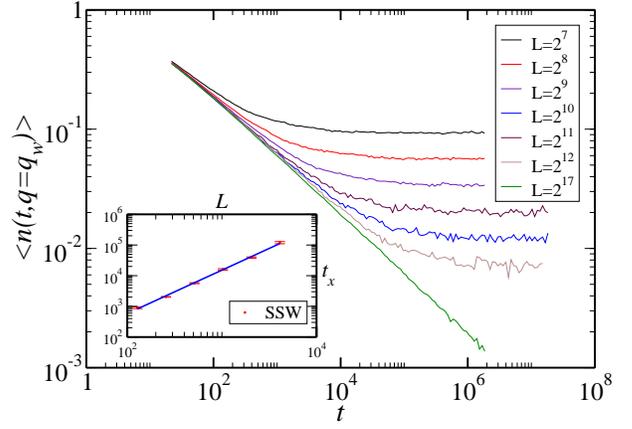}}
\caption{
(Color online)
Main: time decay of the surface order-parameter at $q=q_w$ for
simulations of the SSW model at system-sizes (from top to bottom)
$L=2^7, 2^8, 2^9, 2^{10}, 2^{11}, 2^{12}$, and $2^{17}$
From the lowest curve $\theta_n^{\rm SSW}=0.49(2)$. Inset: the crossover times to saturation as
a function of the system-size lead to $z^{\rm SSW}=1.4(1)$.
The error bars indicate the estimated uncertainties in the crossing point of fits 
of the initial decay and the saturating behavior to straight lines, on the log-log plot.
}
\label{fig3}
\end{figure}

Of more interest is the behavior of the surface order-parameter defined as
$\langle e^{-h} \rangle$ or, equivalently as the density of local contacts
between the interface and the wall. Indeed, considering equation
(\ref{kpz_wall2}) with $D=-\lambda$, the change of variables
$h=-\ln n$ leads to
\begin{equation}
\frac{\partial n(x,t)}{\partial t} =D\nabla^2 n -a_cn -bn^2-cn^3 +n\eta(x,t),
\label{critical_mn1}
\end{equation}
which is a multiplicative noise Langevin equation for the surface
order-parameter \cite{reviews} to be interpreted in the Stratonovich
sense \cite{vankampen}. For sufficiently low values of $b<b_w$ the
interface remains pinned and the stationary density of locally pinned
sites at the wall is high ($n\lesssim 1$). As the transition is
approached (increasing $b$ following path 5 in Fig. \ref{phase_diagram}), the stationary
density of pinned segments goes to zero in a continuous manner as
$\langle n(b,t=\infty) \rangle \sim |b-b_c|^{\beta_n}$. At $b=b_w$,
the interface depins and therefore $\langle h(t) \rangle$ diverges and
$\langle n(t) \rangle$ vanishes with the characteristic exponent
$\langle n(b=b_w,t) \rangle \sim t^{-\theta_n}$.
Obviously, as said before, $\theta_n$ is common to paths 4 and 5 in
figure \ref{phase_diagram}, and the value
$\theta_n=0.5(1)$ has been reported previously \cite{ginelli}.  Here
we focus here on the determination of the path dependent exponents
$\beta_h$, $\beta_n$, and $\nu$. Before proceeding further, we refer
the reader to \cite{haye} for a detailed analysis of the critical
behavior of the above observables in equilibrium critical wetting.

To study numerically the model defined above, owing to well documented
numerical instability problems \cite{problems}, it is more
convenient to integrate numerically Eq. (\ref{critical_mn1}) than 
the equivalent form Eq. (\ref{kpz_wall2}). Equation
(\ref{critical_mn1}) can be efficiently integrated by means of a
recently introduced split-step scheme specifically designed to deal with
Langevin equations with non-additive noise \cite{ivan}. Setting
$D=-\lambda=0.1$, $\sigma=1$ we can use the result $a_c=0.143668(3)$ 
obtained from previous investigations of the analogous nonequilibrium complete-wetting
transition \cite{non-op}. As will be
illustrated below, estimates of the critical exponents are severely
hindered by uncertainties in the value of $a_c$.

To circumvent this problem and to confirm universality we have carried out simulations
of a discrete interfacial growth model which (i) in the absence of
walls is known to belong to the KPZ universality class, (ii) has been
successfully used in nonequilibrium complete wetting analyses
\cite{ssw} and, most importantly, (iii) allows for an exact
determination of the velocity of the free interface, and therefore
permits to extract the critical exponents with good accuracy.
To be more specific, we consider a {\em single step plus wall} model
(SSW) defined as follows.
At time $t$ interface positions above sites $i$ of a one-dimensional 
line of length $L$ are given by integer height variables $h_t(i)$, 
satisfying the {\em solid-on-solid} constrain $|h_t(i)-h_t(i+1)|=1$.
Initially, we take $h_0(2i)=0$ and $h_0(2i+1)=1$. New height configurations 
are generated by choosing at random a site $i$ and growing it to $h_t(i)\to h_t(i) +2$ 
if and only if a local minimum existed at $i$. It can be shown that this rule generates a
KPZ-like interface with $\lambda=-1/2$ moving with an asymptotic
long-time, average velocity $v=(1+L)/(2L)$ \cite{krug_meakin}.
Aditionally, taking advantage of the exact knowledge of $v$, the interface is
globally pulled down by one unit every $L/v$ growth trials, in such a
way that the interface has zero average velocity. 
A wall at $h=0$ is then introduced by precluding the interface from
overtaking the wall that is behind it. This is achieved by
implementing the previous global, downwards movement as $h_t(i) =
|h_t(i)-1|$ for each $i$. Finally, in order to implement an attractive
wall the growing rates at the bottom layer are reduced from $1$ to
$1-q$ with $0\le q \le 1$. The parameter $q$ represents the
short-ranged attraction exerted by the wall on the interface. If
$q=1$ no growth is possible at the local minima located at the wall
and hence an interface at the wall does not move, whereas if $q=0$ the
short-range attraction is switched off.

In all, the SSW algorithm is as follows:
(i) A site $i$ is randomly chosen and grown from $h(i)$ to $h(i) +2$ if a local minimum
exists at $i$ ($h(i+1)+h(i-1)-2h(i)=2$). This is done with probability 1 if $h(i)>0$, 
or with probability $1-q$ if $h(i)=0$, with $0\le q\le 1$.
No action is taken if $i$ does not correspond to a minimum. Time is increased by $1/v$ after $L$
of such attempts.  (ii) Every $L/v$ growth trials $h(i) = |h(i)-1|$
for each $i$. Given that $L/v$ is generally not an integer, this is
done by using $\lfloor L/v \rfloor$ with probability $L/v-\lfloor L/v\rfloor$ and $\lfloor L/v\rfloor+1$ with
probability $\lfloor L/v\rfloor+1-L/v$, where $\lfloor \cdot\rfloor$ denotes the integer part.
Periodic boundary conditions are imposed.
By tuning the parameter $q$ a critical wetting transition is observed. 

\section{NUMERICAL RESULTS}

We next summarize our main findings for both the discrete model (SSW)
and the stochastic differential equation (\ref{critical_mn1}) (SDE).

To determine the critical points, $q_w$ (for the SSW) and $b_w$ (for
the SDE), we take a system-size as large as possible ($L=2^{17}$ here),
plot the order-parameter $\langle n(t) \rangle$ versus $t$ in a
double logarithmic scale, and look for the separatrix between curves
converging to a constant value and those bending downward (not
shown). In this way, the critical values $q_w=0.4445(5)$ (SSW) 
and $b_w=-1.0815(9)$ (SDE) are determined.
In figure \ref{fig3} results are shown in log-log for the time decay of the 
order-parameter $\langle n(t) \rangle$ at the critical point of the SSW model 
for different system-sizes ranging from $L=2^7$ to $L=2^{17}$. From the slope
of a straight-line fit to the lowest curve one finds $\theta_n^{\rm SSW}=0.49(2)$
($\theta_n^{\rm SDE}=0.50(5)$ for the SDE; not shown), where
the error is computed by comparing the slopes corresponding to the upper and
lower bounds of $q_w$ ($b_w$). The inset of Fig. \ref{fig3} 
shows estimates of the crossover times, $t_\times(L)$, from time decay to
saturation as a function of the system-size. Using $t_\times(L)\sim L^{z}$ we find the
temporal exponent values $z^{\rm SSW}=1.4(1)$, $z^{\rm SDE}=1.3(2)$ (not shown), which are in agreement
with those previously reported in \cite{ginelli} 
following path 4 of Fig. \ref{phase_diagram}, and compatible with the theoretical considerations
described above.

\begin{figure}[top]
\centerline{\psfig{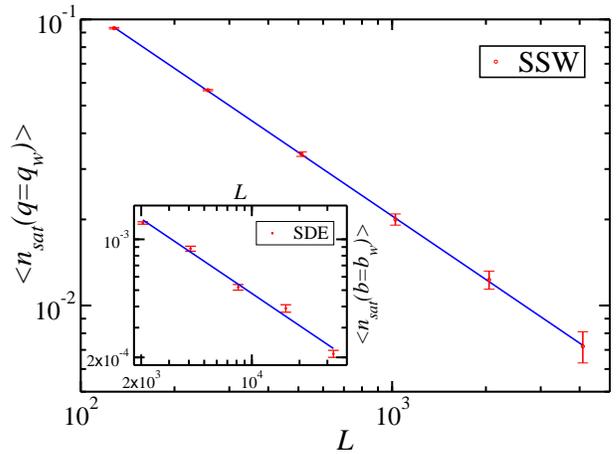}}
\caption{
(Color online)
Main: The SSW saturation values of $\n$ at criticality for several
system-sizes provide the exponent ratio $\beta_n/\nu=0.74(1)$. Inset:
the same analysis for the SDE yields $0.6(1)$. Note the difference in
the magnitude of error-bars, which correspond to three standard 
deviations of the mean saturation values.}
\label{fig4}
\end{figure}

\begin{figure}[top]
\centerline{\psfig{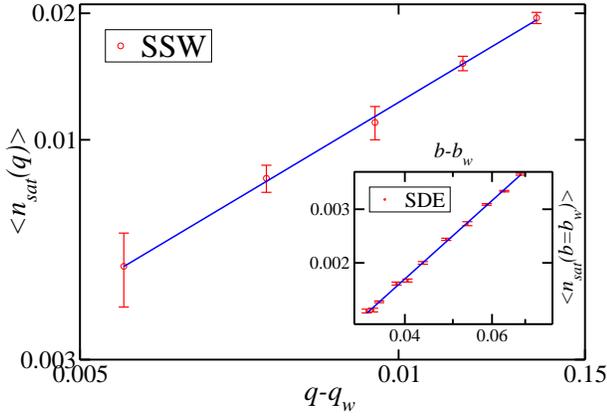}}
\caption{
(Color online)
Main: log-log plot of the saturation values of $\n$ for $L=2^{17}$ 
vs the distance to the critical point gives a direct estimation of
$\beta_n^{\rm SSW}=1.50(9)$ and (inset) $\beta_n^{\rm SDE}=1.46(6)$.
Error bars as in Fig. \ref{fig4}.}
\label{fig5}
\end{figure}

From the scaling of the saturation value at the critical point,
$\langle n_{sat}(q=q_w) \rangle$ for different system-sizes one can
determine (see figure \ref{fig4}) $\beta_n/\nu =0.74(1)$
($\beta_n/\nu=0.6(1)$ for the SDE).
A direct estimation of $\beta_n$ is also possible by measuring the
order-parameter stationary-value for the largest available system-size
upon approaching the critical point from below. This is shown in
figure \ref{fig5} again for both the SSW and the SDE. We find
$\beta_n^{\rm SSW}=1.50(9)$ and $\beta_n^{\rm SDE}=1.46(6)$ which,
along with the obtained values for $\beta_n/\nu$, yields $\nu^{\rm
SSW}=2.0(2)$ and $\nu^{\rm SDE}=2.4(5)$ (note the relatively large
errorbar in this latter case). These values supersede the early
estimate $\beta_n=1.2$ given in \cite{nos}

\begin{figure}[top]
\centerline{\psfig{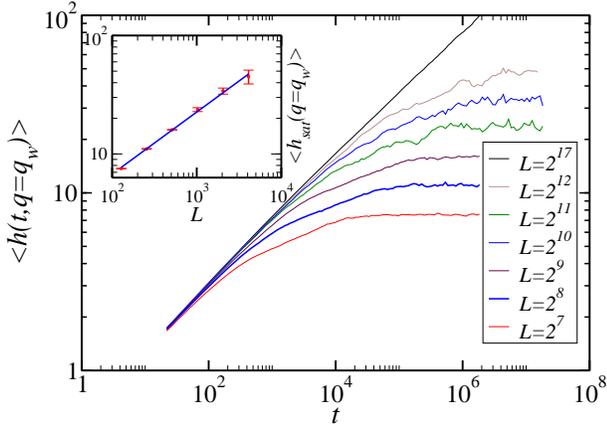}}
\caption{
(Color online)
Main: log-log plot of the average distance to the wall vs time,
yielding $\theta_h^{\rm SSW}=0.35(2)$. System-sizes are (from top to bottom)
$L=2^{17}, 2^{12}, 2^{11}, 2^{10}, 2^{9}, 2^{8}$, and $2^7$.
Inset: finite-size scaling of 
$\langle h_{sat}(q=q_w) \rangle$ indicating
$\beta_h/\nu=0.52(4)$. All values are for the SSW model.
Error bars as in Fig. \ref{fig4}.}
\label{fig6}
\end{figure}

\begin{figure}[top]
\centerline{\psfig{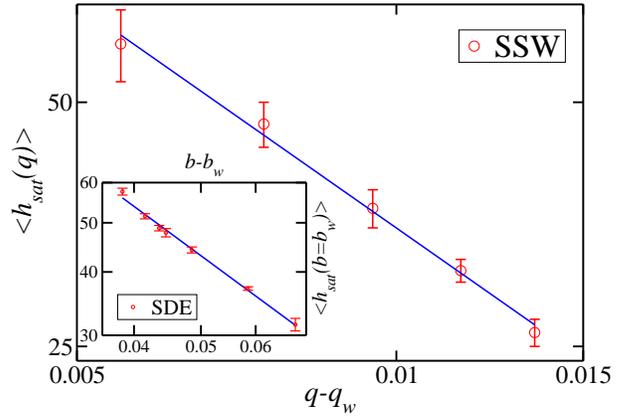}}
\caption{
(Color online)
Main: the scaling of the saturation value of $\langle h_{sat} \rangle$ yields
$\beta_h^{\rm SSW}=0.9(1)$ for data collected from the SSW model. Inset:
A value $\beta_h^{\rm SDE}=1.0(1)$ results for data collected from the SDE.
Error bars as in Fig. \ref{fig4}.}
\label{fig7}
\end{figure}

The scaling properties of the mean interfacial separation $\h$ can be
determined analogously (we show results only for the SSW model).
The time growth of $\langle h(t) \rangle$ for the largest system available $L=2^{17}$
yields $\theta_h^{\rm SSW}=0.35(2)$ (see Fig. \ref{fig6}), in reasonable agreement with
the expected value $\theta_h=1/3$.  The average saturation values
are plotted in log-log as a function of the system-size in the inset of
figure \ref{fig6}). From them we estimate $\beta_h/\nu=0.52(4)$ (compatible with
KPZ scaling, as $\zeta=\beta_h / \nu =1/2$). Estimations of $z$ can be
analogously obtained, but they are rather noisy because of the uncertainty 
in determining the saturation value.  Also, a direct estimation of
$\beta_h$ can be obtained for the largest available size. This leads
to $\beta_h^{\rm SSW}=0.9(1)$ and $\beta_h^{\rm SDE}=1.0(1)$ (see figure
\ref{fig7}). Using the values of $\beta_h/\nu$ and $\beta_h$ a third estimate
for $\nu=1.8(3)$ is obtained.

Finally, we have confirmed that for both models at the tricritical
point the width, $W$, of the interface grows with time with an
exponent compatible with that of the KPZ, $W(t) \sim t^{\theta_W}$, with
$\theta_W=1/3$, and saturates in finite system-sizes to a
non-vanishing value given by $W \sim L^{-\beta_h/\nu} \sim L^{-1/2}$
as in the KPZ.

In summary, the exponent values determined above appear to be
compatible with the set of simple rational numbers
$\theta_n=1/2, z=3/2, \beta_n=3/2, \nu=2, \theta_h=1/3$, and  $\beta_h=1$.
For the sake of comparison, the critical exponents at equilibrium
critical wetting are 
$\theta_n=1/4, z=2, \beta_n=1, \nu=2, \theta_h=1/4$ and $\beta_h=1$
\cite{shick}.  The coincidence in the numerical
values of two of them (the path dependent ones, $\beta_h$ and $\nu$)
is somehow intriguing.

\section{THEORETICAL CONSIDERATIONS}

It is instructive to compare the above results with those obtained at
high system dimensionalities from a self-consistent, mean-field
approximation to equation (\ref{critical_mn1}) \cite{cw_meanfield}. In
that approximation, a sequence of three scaling regimes was reported
to exist depending on the relative importance of the noise strength as
compared to the spatial coupling. The first two have a Gaussian
character, while in the third regime, i.e. the strong noise one (in
which the wetting temperature is shifted away from zero) all moments,
$m_k=\langle n^k \rangle$, for $ k > 2D/\sigma^2 +1$ scale with the
same exponent, while simple scaling $m_k \sim m^k$ is obtained for
$k\le 2D/\sigma^2+1$. This is a rather curious type of anomalous
scaling not very different from that observed in analogous mean-field
approximations for non-equilibrium complete wetting \cite{Colaiori}.
However, we have verified numerically that in the one-dimensional
system moments of arbitrary order scale as $\n$ itself, as happens in
other Langevin equations with multiplicative noise \cite{reviews}.
This is a consequence of the large fluctuations occurring at $d=1$,
notwithstanding which it is possible to make some analytic
predictions, as we now discuss.

First, we explicitly show how the critical value of $b$ is depressed
from its mean-field value $b_w=0$ to $b_w < 0$ once fluctuations are
included.  Taking spatial and noise averages in the stationary state
of equation (\ref{kpz_wall2}) for a generic value of $a$, and denoting
the average squared slope of the interface by $s^2\equiv
\langle(\nabla h)^2\rangle$, we obtain
\begin{equation}
a-s^2 +b\n +\langle n^2 \rangle=0
\label{eqn}
\end{equation}
where, without loss of generality we have taken $D=\lambda=-1$ and
$c=1$. At coexistence $a_c=s^2_c$, and hence $s_c^2-s^2+b\n+\langle
n^2 \rangle=0$. Additionally, in the pinned phase (i.e. $\n\neq 0$)
$s^2<s^2_c$ and therefore equation (\ref{eqn}) has a solution only if
$b<0$. Still, $b_w$ could be $0$, but by noting that on approaching
the tricritical point along path 5 the moments of $\n$ scale in the
same way, $\n \sim \langle n^2 \rangle \sim |b-b_w|^{\beta_n}$, it is
easy to see that $b_w \neq 0$ because otherwise the negative term
$b\n$ would be subdominant in comparison with the positive $\langle
n^2 \rangle$, and no solution could exist for small $b$. Clearly, by
continuity $b_w$ cannot be positive and hence $b_w<0$ ensues.

It is illuminating to show how a similar reasoning leads to a more
predictive analysis when applied to the analogous complete wetting
transition (path 1 in Fig. \ref{phase_diagram}) known to be in the MN1 class. In that
case, the condition $a-a_c + s_c^2-s^2 +b\n=0$ must be satisfied, with
the negative term $\delta a=a-a_c$ balancing the two positive terms
$s_c^2-s^2$ and $b\n \sim |\delta a|^{\beta_n}$ in the pinned
phase. This implies that $\beta_n>1$ if a solution for $\n$ is to
exist for small $\delta a$ \cite{grinstein}.  Furthermore,
by noticing that if KPZ scaling is applicable, then
$s_c^2-s^2 \sim \xi^{2(1-\zeta)}$ \cite{krug}, where $\zeta$ is the roughness 
exponent of a free KPZ. Recalling that $\zeta=1/2$ in 1$d$,
immediately entails $\nu=1$ for complete wetting \cite{grinstein}.

On the contrary, we have not been able to derive the value of $\nu$
for nonequilibrium critical wetting, nor does it seem immediate that $\nu=1$
along path 4 as found numerically in \cite{ginelli}.  The main
difference with the complete wetting case appears to rest on the
behavior of the slopes $s^2$.  According to our measurements for
equation (\ref{kpz_wall2}), $s_c^2-s^2 \sim |b-b_w|^{1.51(2)}$ for
critical wetting (path 5) and $s_c^2-s^2 \sim |a-a_c|^{0.74(2)}$ for
tricritical complete wetting (path 4), pointing to an anomalous
scaling of the slopes, i.e. $s^2 = \langle (\nabla h)^2 \rangle$ does
not have the same scaling dimension as $[L^{-1}]^2 [h]^2$, as happens
in the complete wetting case.

That $s_c^2-s^2$ scales along path 4 with an exponent less than unity
could have been anticipated from the condition $|\delta a|-b_w |\delta
a|^{\beta_n} =s_c^2-s^2$ at the tricritical point, implying $s_c^2-s^2
\sim |\delta a|^{\beta_s}$ with $\beta_s \leq 1$.

The exponent value $\beta_n=0.74(5)$ reported in \cite{ginelli} for
the multicritical complete wetting transition along path 4 (confirmed
in our own measurements), along with $\beta_n \approx 1.5$ for
critical wetting as referred above, indicates that the anomalous
scaling of the slopes is ultimately controlled by $\beta_n$ in both
cases, rather than by the roughness exponent of the KPZ as in the
complete wetting case. Notice the constancy along path 4 and 5 of the
ratio $\beta_n/\nu \approx 0.75$, which is a property of the
tricritical point and therefore must be independent of the path.

The fact that the slopes acquire a scaling not directly derivable from
the free KPZ can be argued to be a consequence of the effect of the
potential well on the interface.  Figure \ref{fig8} shows snapshots of
configurations $h(x)$ that result from solving the SDE in the
stationary state for different parameters that correspond to approaching 
the critical points from different paths: 

\begin{itemize}
\item Panel A for $(b=-1.2<b_w,a=-0.25<a_c)$; corresponds a situation slightly
below the complete wetting (MN1) transition (path 1 of
Fig. \ref{phase_diagram}).

\item Panel B for $(b=b_w, a=-0.153<a_c)$, slightly below the 
 multicritical complete wetting transition (path 4).

\item Panel C for $(b=-1.3<b_w, a=a_c)$, slightly below the
 critical wetting transition (path 5).
\end{itemize}

The distances to the transition points are chosen so that $\h \approx
5$ in all cases. Note the clear qualitative difference between panel
A, for which standard scaling holds, and the rest. Observe that in
panel C, the interface consists of patches of essentially free KPZ
interfaces separated by regions of sites pinned by the potential
well. It then seems plausible to conclude that, following path 5,
regions locally trapped within the potential well develop slopes
different from that of a free KPZ (indeed, roughness is severely
restricted within the potential well) even at points arbitrarily close
to the tricritical point, thereby inducing an anomalous scaling
controlled by $\beta_n$.  In panel B, i.e. upon approaching the
tricritical point along path 4, the effect of the bounding wall is
less apparent as the potential well is marginally disappearing at
$b=b_w$, but a similar effect, induced by the potential shape, should
be at work in this borderline case.

\section{SUMMARY AND DISCUSSION} 

We have investigated the universal properties of nonequilibrium
critical wetting transitions in one spatial dimension. For that, we
study effective interfacial models in the KPZ universality class with
$\lambda <0$ bounded by a lower wall, and determine the average height
$\langle h \rangle$ and the surface order-parameter $\n=\langle e^{-h}
\rangle$.

A scaling analysis leads to the prediction $z=3/2$ and a time behavior
of the average height governed by the growth exponent of the free KPZ,
i.e. $\langle h(t) \rangle \sim t^{1/3}$. These results have been
verified numerically. Other exponents have been computed from
extensive numerical simulations of the Langevin equation and a
discrete model in the same universality class that enables a more precise numerical analysis,
as a result of which we find $\nu \simeq 2$, $\beta_h\simeq 1$, and $\beta_n\simeq
3/2$, suggesting that actually the exponents take rational values
(see table I). Interestingly enough, the first two values agree with
those of equilibrium critical wetting. 

Simple analytical arguments allow us to show that the critical value
of the control parameter $b$ is depressed by fluctuations from its
mean-field value $b_w=0$ to $b_w <0$. We have also shown that in
critical wetting, as well as in multicritical complete wetting, the
average interface slopes do show anomalous scaling, not controlled by
the free KPZ equation: local regions pinned by the binding potential
generate anomalous scaling.

We have not been able to predict the value of $\nu$ using the same
analytical considerations that yield $\nu=1$ in the complete wetting
(MN1) case, nor is it trivial to obtain $\nu=1$ for multicritical
complete wetting.  It is nevertheless possible to conclude that in
this latter case the exponent governing the scaling of the interface
slopes is less than unity, $s_c^2-s^2
\sim |a-a_c|^{\beta_s}$, in agreement with the value $\beta_s \approx 0.75$
obtained. There is numeric evidence that the slopes actually scale
with $\beta_s=\beta_n$ for both critical wetting and multicritical
complete wetting (paths 5 and 4 of Fig. \ref{phase_diagram},
respectively), a result that violates na\"ive scaling. In effect,
dimensional analysis demands that $s_c^2-s^2$ scales as
$|a-a_c|^{2(1-\zeta)}$, $\zeta$ being the roughness exponent of the
free KPZ \cite{krug}. This is obeyed at the complete wetting
transition, and indeed was used to derive $\nu=1$ in MN1
\cite{grinstein}, but does not hold for critical wetting nor for
multicritical complete wetting.

\begin{figure}[top]
\centerline{\psfig{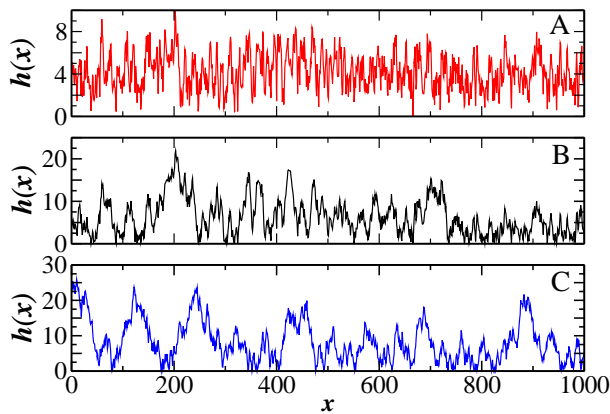}}
\caption{
(Color online)
Snapshots of configurations $h(x)$ as results from solving the SDE in
the stationary state for (A) $b=-1.2<b_w, a=-0.25<a_c$ (complete
wetting), (B) $b=b_w, a=-0.153<a_c$ (multicritical complete wetting),
and (C) $b=-1.3, a=a_c$ (critical wetting).  The distances to the
transition points are chosen so that $\h \approx 5$ in all cases. In
panels (B) and (C) patches of depinned interfaces are observed where
$n\approx 0$.}
\label{fig8}
\end{figure}

The cause of the deviation from standard scaling can be sought in the
interfacial profiles shown in figure \ref{fig8}. For critical and
multicritical wetting, patches of free KPZ interfaces ($n(x) \approx
0$) are separated by regions of sites that lie in the potential well,
plausibly hindering the standard KPZ scaling from setting in gradually
as the tricritical point is approached. This is in contrast to typical
interfacial profiles for nonequilibrium complete wetting (MN1) where
no potential well is at play.

\begin{table}[top]
\caption{Summary of the critical exponents for nonequilibrium,
critical wetting transitions with short-range forces. Results are
shown for the discrete model SSW, the stochastic differential equation
SDE, and from other sources when available.\label{table}}
\centerline{\psfig{figure=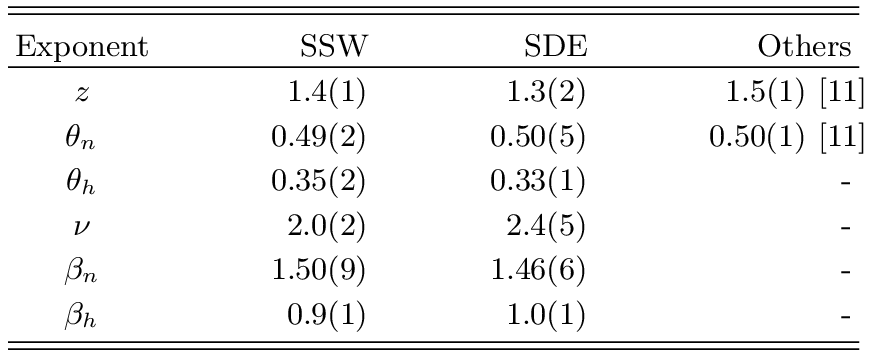,angle=0,width=8.0cm}}
\end{table}

According to a recent self-consistent, mean-field approximation to
equation (\ref{critical_mn1}) three different scaling regimes of critical
behavior for the surface order-parameter can be distinguished
\cite{cw_meanfield}. The first two are of Gaussian type, while the
third one is a highly nontrivial strong-fluctuating regime.  This rich
structure is completely washed out by fluctuations in one-dimensional
systems, where a unique and universal scaling regime emerges.
Moreover, we have verified numerically that in the one-dimensional
system moments of arbitrary order scale as $\n$ itself, as happens in
other Langevin equations with multiplicative noise \cite{reviews}.
The fact that the numerical values of the exponents are changed is not
surprising at all, given the presence of severe fluctuations in
one-dimension but what is more striking is that out of the three
regimes appearing in the mean-field approach, only one survives.  A
possible explanation for such an abrupt change might come from a
recent claim that the strong-coupling renormalization group fixed
point of the KPZ dynamics is essentially different above and below
$d=2$ \cite{Leonie}.  This point remains to be further studied, as
well as some other aspects of nonequilibrium critical wetting
including its subtle relation to its equilibrium counterpart (signaled
by the coincidence of some exponents), and the possible existence of
various nonuniversal scaling regimes in higher dimensions. It also
stands as a main experimental challenge to observe in the laboratory
the phenomenology reported on here and in previous theoretical works
of nonequilibrium wetting.

\begin{acknowledgments}
One of us (O.A.H) would like to thank F. Ginelli for fruitful discussions.
This work was supported in part by the
Spanish projects FIS2005-00791 and FIS2005-00973 (Ministerio de Ciencia y
Tecnolog\'ia), FQM-165 and FQM-0207 (Junta de Andaluc\'ia). 
\end{acknowledgments}

\end{document}